# Tunable quantum two-photon interference with reconfigurable metasurfaces using phase-change materials

NOOSHIN M. ESTAKHRI[1,2*] AND THEODORE B. NORRIS,[1,2]

[1]*Department of Electrical Engineering and Computer Science, University of Michigan, Ann Arbor, Michigan 48109, USA*
[2]*The Gérard Mourou Center for Ultrafast Optical Science, University of Michigan, Ann Arbor, Michigan 48109, USA*
*\*estakhri@umich.edu*

**Abstract:** The ability of phase-change materials to reversibly and rapidly switch between two stable phases has driven their use in a number of applications such as data storage and optical modulators. Incorporating such materials into metasurfaces enables new approaches to the control of optical fields. In this article we present the design of novel switchable metasurfaces that enable the control of the nonclassical two-photon quantum interference. These structures require no static power consumption, operate at room temperature, and have high switching speed. For the first adaptive metasurface presented in this article, tunable nonclassical two-photon interference from -97.7% (anti-coalescence) to 75.48% (coalescence) is predicted. For the second adaptive geometry, the quantum interference switches from -59.42% (anti-coalescence) to 86.09% (coalescence) upon a thermally driven crystallographic phase transition. The development of compact and rapidly controllable quantum devices is opening up promising paths to brand-new quantum applications as well as the possibility of improving free space quantum logic gates, linear-optics bell experiments, and quantum phase estimation systems.

## 1. Introduction

Two-photon interference is an example of a pure quantum mechanical effect with no classical analog and is one of the fundamental phenomena underlying the field of quantum optics. This effect was first observed in Hong-Ou-Mandel (HOM) fourth-order interferometry [1] in which two indistinguishable photons are superposed at a lossless beam splitter, entering two separate input ports, and exiting together in either one or the other output port (bunching).

This interference technique has often been used in dual-arm or common-path geometries for measuring the indistinguishability of the photons generated in a spontaneous parametric down-conversion process (SPDC) [2] or the consecutive photons emitted from single-photon sources such as semiconductor quantum dots [3], or testing the bosonic nature of surface plasmon polaritons [4]. Two-photon interferometry via HOM and similar methods have also been used to measure the single photon tunneling time across a barrier of 1D photonic band-gap material [5], to find the polarization-mode dispersion in birefringent materials [6], and to simultaneously determine the group delay and phase delay imposed on orthogonally polarized photons [7]. Recently, the temporal resolution of dual-arm HOM interferometry has been improved to a few attoseconds [8], expanding the range of its capability towards studying sub-nanometer thick samples.

Two-photon interference is the basis for several applications in quantum information sciences, such as realization of quantum logic gates [9], generation of multiphoton

Greenberger-Horne-Zeilinger entanglement [10], linear-optics Bell measurements [11], boson sampling [12,13], and quantum information processing circuits [14].

Applications requiring multi-photon interference conventionally employ bulk 3D optical beam splitting components, hindering our ability to make compact quantum optical systems. These realizations are also often static in the sense that the electromagnetic properties of the setup components are fixed. Although a static setup is sufficient for some applications, it would be highly desirable to have switchable or adaptive components that can deliver *tunable* two-photon interference. For example, in quantum applications that inherently require alterations in the quantum setups, such as quantum state tomography [15], physical movement or modification of components introduces new sources of error, lengthens the measurement process, and degrades the overall performance.

This paper addresses the need in quantum optics for free-space compact devices creating tunable quantum interference and acting as fast modulators for second order intensity quantum correlations. Specifically, we present a solution to this challenge using switchable metasurfaces incorporating a phase-change material (PCM), germanium telluride (GeTe). Our metasurface-based geometries are 2x2 coupling elements, operating as an extension of traditional beam splitters (see supplemental document). The incorporation of tunable lossy material enables switching between constructive and destructive two-photon quantum interference. The proposed devices consume zero static power, as the phase-change material (GeTe) is stable in both crystalline and amorphous phases. Unlike cases of reconfigurable quantum photonic chips that use voltage-controlled thermal phase shifters [16], our structures operate in free space. The ultra-thin reconfigurable metasurface-based geometries developed in this paper are a step towards compact, tunable, and robust free-space quantum devices and circuits with high modulation speed and zero-static power consumption.

Optical metamaterials and metasurfaces, both passive and active, have been extensively developed for an enormous range of applications in the control of classical optical fields. In a similar fashion, efforts have begun to utilize metasurfaces for control of quantum light, both for achieving new capabilities and for enabling compact quantum optical systems. For example, an all-dielectric passive metasurface geometry operating in transmission has been used to construct nonclassical multiphoton interferences from polarization encoded states [17]. In the device structures explored here, a different structure of the input and output ports is employed, and the nonclassical two-photon interference is controlled between co-polarized input photonic fields. In ref [17], an all-dielectric metasurface-based implementation is utilized to avoid plasmonic losses; here, however, we intentionally harness the presence of controllable loss channels in designing nonclassical interference [18] and thus enable robust tunable two-photon interference in free space.

To enable active modulation of the quantum interference, we incorporate a phase-change material into the metasurface-based 2x2 networks. PCMs are widely used in classical nano-photonics platforms such as photonic integrated circuits and metasurfaces to achieve reconfigurable functionalities. These are materials with crystalline and amorphous phases that can be electrically or thermally switched between these structurally distinct phases. PCMs exhibit pronounced contrast in the optical properties between the two phases along with other properties such as reversibility, fast switching speeds in orders of ten to a few hundred nanoseconds [19-21], and several-decades-long stability in one phase. Applications include rewritable data storages (such as DVDs) [22,23], on-chip photonic memories [24], and solid-state display technologies [25]. GeTe compound is the PCM used in this paper to add tunability to the design; details of the material are given in sections 2.2, 3, and supplemental document.

In the following, we start by reviewing the joint probability for the detection of photons in the output ports of a general 2x2 network (including loss) and the constraints on general passive networks. We then propose two adaptive planar metamaterial geometries to optimize the tunability for nonclassical two-photon interference in each geometry. Several simulation results are discussed including a study on the effect of partial crystallization of the phase-change material. Next, the temporal form of HOM interference is obtained for each geometry followed by a heat transfer analysis exploring the joule heating scheme in the adaptive planar structures.

## 2. Quantum interference with ultra-compact structures

In this section we review the properties and constraints on the two-photon interference phenomenon in general passive networks (section 2.1), and present two adaptive planar metamaterial geometries to realize tunable constructive and destructive nonclassical two-photon interference (section 2.2).

*2.1 2x2 lossless and lossy networks, photon probabilities, and energy constraints*

The schematic for a universal 2x2 network coupling two input modes with bosonic or fermionic nature is shown in Fig. 1(c). The same representation is useful for modeling the interactions between two incident classical electromagnetic fields via incorporating cross-coupling between the two input ports of the network. This 2x2 network is employed to explore the tunability of two-photon interference phenomena, centered on the interference of two-photon probability amplitudes in a two-photon detection scheme. The continuum annihilation operators of the quantized electromagnetic fields for the two independent input (output) modes at ports $a$ and $b$ are $\hat{a}_{in}(\omega)$ and $\hat{b}_{in}(\omega)$ ($\hat{a}_{out}(\omega)$ and $\hat{b}_{out}(\omega)$) in second quantization. The 2x2 complex transmission matrix $T$ represents the relation between the two input and output classical electromagnetic modes. In the presence of loss, Langevin noise operators, $\hat{F}_a(\omega)$ and $\hat{F}_b(\omega)$, associated with fluctuating currents in the presence of loss, must be included in the relation between the input and output annihilation operators as written in Eq. (1).

$$\begin{pmatrix} \hat{a}_{out}(\omega) \\ \hat{b}_{out}(\omega) \end{pmatrix} = T \begin{pmatrix} \hat{a}_{in}(\omega) \\ \hat{b}_{in}(\omega) \end{pmatrix} + \begin{pmatrix} \hat{F}_a(\omega) \\ \hat{F}_b(\omega) \end{pmatrix} \quad , \quad T = \begin{pmatrix} \tilde{t}_1 & \tilde{t}_2 \\ \tilde{t}_3 & \tilde{t}_4 \end{pmatrix}. \tag{1}$$

Note that these Langevin noise operators must be incorporated in such a way to maintain the validity of the commutation relations between the creation and annihilation operators at the input and output ports (Eq. (2)). These equations produce the commutation relations between noise operators as listed in [26,27].

$$\begin{aligned} \left[\hat{a}_j(\omega), \hat{a}_j^\dagger(\omega')\right] &= \left[\hat{b}_j(\omega), \hat{b}_j^\dagger(\omega')\right] = \delta(\omega - \omega') \\ \left[\hat{a}_j(\omega), \hat{b}_j^\dagger(\omega')\right] &= \left[\hat{b}_j(\omega), \hat{a}_j^\dagger(\omega')\right] = 0 \end{aligned} \quad \forall j \in \{in, out\} \tag{2}$$

We consider the state at the input of the network as a two-photon Fock state of the form in Eq. (3), modeling the outcome of a spontaneous parametric down-conversion process with the normalized two-photon spectrum amplitude $\psi(\omega_a, \omega_b)$ accounting for the characteristic frequency spreads of the down-converted signal and idler photons.

$$|\psi\rangle = \int_0^\infty d\omega_a \int_0^\infty d\omega_b \psi(\omega_a, \omega_b) \hat{a}_{in}^\dagger(\omega_a) \hat{b}_{in}^\dagger(\omega_b) |0\rangle \tag{3}$$

For this input state, the joint probability of the detection of photons at two detectors at the output of the network utilizing the Kelley-Kleiner counting formalism [28,29], and for a relatively large coincidence counting window is as follows [18]:

$$P(1_a, 1_b) = K \langle \hat{N}_a \hat{N}_b \rangle = K \left\{ |\tilde{t}_2|^2 |\tilde{t}_3|^2 + |\tilde{t}_1|^2 |\tilde{t}_4|^2 + 2\operatorname{Re}\left\{\tilde{t}_1 \tilde{t}_4 \tilde{t}_2^* \tilde{t}_3^*\right\} I_{overlap} \right\} \tag{4}$$

Here the nonideal detector efficiencies are incorporated in the constant coefficient $K$ and the continuum number operators with linear superposition over spectrum [30] in the output ports $a$ and $b$ are $\hat{N}_a$ and $\hat{N}_b$ respectively. The HOM two-photon interference [1] for conventional 50:50 beam splitters and the resulting coalescence (bunching) can be seen from Eq. (4). In that case, the probability at relative time zero drops to the value of zero for unity overlap integral. The quantum overlap integral and the baseline of the joint probability, calculated for distinguishable photons in the input ports are:

$$I_{overlap} = \int_0^\infty d\omega \int_0^\infty d\omega' \psi(\omega,\omega')\psi^*(\omega',\omega) \quad , \quad Baseline = |\tilde{t}_2|^2|\tilde{t}_3|^2 + |\tilde{t}_1|^2|\tilde{t}_4|^2 \tag{5}$$

Note that the phase dependence shows up only in the quantum interference term of Eq. (4); hence, the total phase of the network (i.e. $\angle \tilde{t}_2 + \angle \tilde{t}_3 - \angle \tilde{t}_1 - \angle \tilde{t}_4$) can be utilized to control the fourth order interference [18]. With proper choice for the reference planes of the input and output ports of the network, whether in free space or electromagnetic waveguide geometries, one may push all the phase information to the off-diagonal elements of the transmission matrix and keep the $\tilde{t}_1$ and $\tilde{t}_4$ elements purely real. For a passive network, which may be asymmetric or unbalanced, the total output energy of the network must always be equal or less than the input energy. This leads to an energy constraint inequality [26] implying that, for lossless 2x2 networks, regardless of the amplitudes of the transmission matrix elements, the total phase is always equal to $\pi$. An immediate consequence from Eq. (4) is that for lossless 2x2 systems the HOM experiment can only result in a *destructive* interference for the two-photon input state. Hence, loss needs to be added to the design, as we do here, to enable *constructive* two-photon quantum interference [18].

*2.2 Metasurface structures for tunable quantum interference*

Metasurfaces, the flat optical counterparts of metamaterials, are built from array(s) of scattering elements. These elements are engineered to locally manipulate incident electromagnetic fields with the purpose of inducing a broad range of functionalities; familiar examples include dispersion engineering and light wave focusing via flat optics.

More recently, spatiotemporally varying metasurfaces have been explored. Such active metasurface-based devices have offered new possibilities [31,32]. The mechanisms for tuning may be electrical [33-35], optical [36-41], magnetic [42], mechanical [43-45], thermal [46,47], or chemical [48,49]—such as hydrogenation. Tunable metasurface-based devices have been demonstrated using reconfigurable platforms such as materials demonstrating ferroelectricity below the phase transition temperature [35,50] or after being transformed into their paraelectric state; i.e. above the Curie temperature [46]. Other tunable or switchable materials such as superconducting films [51,52], phase-change materials [53-55], liquid crystals [42,47], and graphene [56] may also be employed to realize adaptive metasurfaces.

In this work we focus on designs utilizing electric stimuli for the primary reason that fast tunability may be obtained, while many other approaches are intrinsically slow. Specifically, we are interested in phase-change materials. A number of materials with different phases such as Ge2Sb2Te5 (GST) [53], Ge3Sb2Te6 (GST) [54] and Vanadium dioxide (VO2) [55] have been recently used to achieve dynamic control; here we utilize GeTe, a chalcogenide compound with two stable phases, in the adaptive metasurface design. Details regarding this material may be found in section 3 of this paper and the supplemental document.

Figs. 1(a) and (b) illustrate two planar adaptive metasurface structures designed to implement tunable quantum interference of two-photon state. The first geometry (structure A) is a patterned structure composed of alternating GeTe and gold (Au) nano-strips resting on a SiO2 substrate. The gold strips function as an integrated nano-heater, which can be deposited using E-beam lithography, with GeTe being subsequently sputtered on the structure. The second geometry (structure B) is a layered structure composed of fused-silica ($SiO_2$), germanium telluride (GeTe), and titanium dioxide ($TiO_2$) layers resting on a layer of Titanium Nitride (TiN) as the heating sheet. The choice of the heating layer in this structure is based on the advantages of TiN thin films such as their thermal stability, higher melting point, low cost, as well as compatibility with the standard silicon manufacturing processes [57,58] in comparison to other possible metallic layers such as gold and silver.

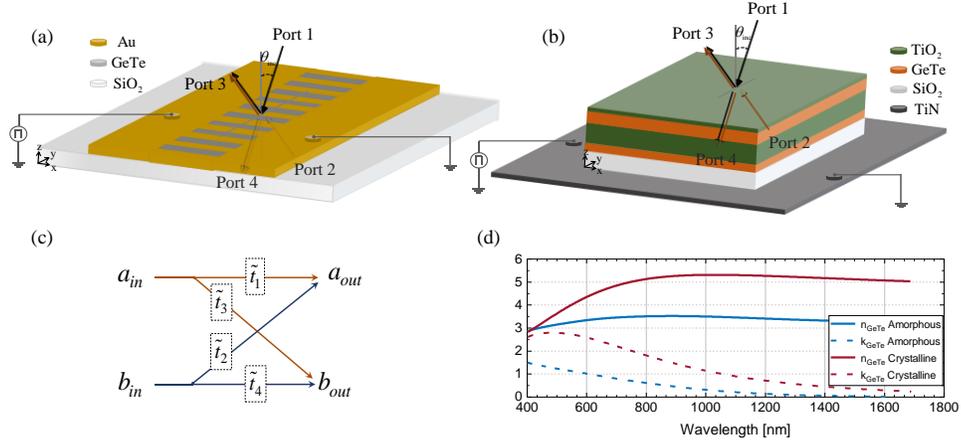

Fig. 1. Schematic illustration of two metasurface structures composed of (a) alternating GeTe and gold nano-strips with rectangular cross sections on a fused-silica substrate with gold strips also used for heating purposes (structure A), (b) SiO2, GeTe, and TiO2 layers resting on a TiN heating film (structure B). (c) General 2x2 network with $\hat{a}$ and $\hat{b}$ representing the annihilation operators of quantized electromagnetic fields at the input and output. (d) Real and imaginary parts of the refractive indices of GeTe at crystalline and amorphous phases.

Both structures are designed to perform as 2x2 network arrangements with port numberings and the direction of input and output ports shown schematically in Fig. 1(a) and (b). Here, all the port angles, which can be chosen freely, are set at 45 degrees. In the patterned periodic geometry of Fig. 1(a) the unit cell width is set at 450 nm in order to support only the fundamental zeroth order Floquet mode and block all the higher order modes at the operating wavelength of 810 nm. This wavelength is common for correlated photons generated via SPDC in BBO crystals. For the chosen conditions, all higher order modes are blocked for unit cell widths less than 474 nm.

Electromagnetic full wave analysis [59] is employed to simulate both adaptive metasurface geometries to obtain the scattering matrix elements. Input ports are excited with transverse electric (TE) polarization. The designs are optimized with the goal of maximizing the degree of two-photon interference tunability in each structure while setting a lower limit on the baseline of the two-photon interference for distinguishable photons. This minimum acceptable baseline is set at 0.0833 (=1/12) and 0.0625 (=1/16) for the structures A and B, respectively, and is chosen to avoid unphysical and extremely lossy final configurations. (Parametric studies of the designs are given in section 3 followed by discussions regarding wavelength tuning and bandwidths.) The final patterned adaptive metasurface geometry has the following parameters: the unit cell width is set at 450 nm with the thickness of 190 nm and 15 nm for $SiO_2$ and GeTe/Au layers, respectively. Also, the GeTe strip width in the final configuration is set at 285 nm. For the adaptive layered geometry, the thickness of TiN, SiO2, lower GeTe, lower $TiO_2$, upper GeTe, and upper TiO$_2$ are optimized at 15 nm, 290 nm, 21 nm, 330 nm, 13 nm, and 290 nm, respectively.

## 3. Tunable nonclassical two-photon interference: analysis and results

GeTe is a typical PCM such as Ge$_2$Sb$_2$Te$_5$, Ge$_1$Sb$_2$Te$_4$, and Sb$_2$Te$_3$ that lies on the edge of the pseudobinary line GeTe-Sb$_2$Te$_3$. It is a chalcogenide compound with two stable phases, crystalline and amorphous, and the phase can be switched by changing the temperature [60]. The phase change can be induced fast and repeatedly through local heating process using electrical or laser heating pulses as short as ten to several hundred nanoseconds [19-21]. In switching from the amorphous phase to crystalline and then back, the first transition is more time-consuming and hence determines the overall switching speed of the device. To determine the limits to the fastest possible speeds, a time-resolved resistance analysis of a sample classical bottom-heater GeTe geometry demonstrates that SET pulses as short as 16 ns may be utilized to successfully crystalize these materials [20]. Depending on the resistance of the RESET mode, under a set of conditions, a SET pulse as short as a couple of nanoseconds may even suffice [20]. We note that the time domain analysis presented in the final section of this article is not intended to study the behavior of the device at the highest possible speeds, but with the purpose of analyzing our device geometries at slightly lower yet more common and manageable speeds [21].

As GeTe changes its phase from amorphous to crystalline, which typically happens rapidly within a few degrees Kelvin, the electrical resistivity of GeTe changes by several orders of magnitude. As an example, reported by [61], the electrical resistivity of a $2250 \overset{\circ}{A}$ GeTe film decreases from $10^3$ to $10^{-4} [\Omega.cm]$ in transition from amorphous to crystalline phase measured at room temperature. The transformation temperature is almost fixed for GeTe layers of thickness greater than 350 nm, with only a small dependence on the thin film deposition temperature or other deposition parameters [61]. In a recent observation reported in [62], the GeTe sheet resistance (240 nm thick films) shows a sharp decrease (also 7 orders of magnitude) around the measured annealing temperatures of $150^\circ C$ as the phase transition happens, with a slight dependence of the transition temperature on the annealing temperature. The binary composition employed in this article, GeTe, like other PCMs on the pseudobinary GeTe-Sb$_2$Te$_3$ line, has two different crystalline phases: rock-salt structure and trigonal structure [60] (See supplemental document for discussions with regard to the transition of GeTe back and forth between amorphous and crystalline phases.) The transformation of the in-built geometry of the GeTe compound in these transitions results in a significant change in the value of the material refractive indices. Fig. 1(d) shows real and imaginary parts of the refractive indices for a 200 nm film of GeTe in crystalline and amorphous phases [adapted from 21].

### 3.1 Coalescence and anti-coalescence

Eq. (4) links the amplitudes and phases of the scattering parameters to the coincidence counts in the output ports, which is proportional to the probability of having a single photon in each output. To quantify the nonclassical two-photon interference, we calculate and plot the Hong-Ou-Mandel (anti-) coalescence for each metasurface configuration utilizing the scattering parameters. Coalescence (anti-coalescence) is the ratio of the decrease (increase) in the coincidence counts at relative delay time zero due to two-photon interference to the value of the baseline coincidence counts which corresponds to distinguishable inputs. Therefore, from Eqs. (4) and (5) one may determine (anti-) coalescence parameter as:

$$\text{(Anti-)Coalescence} = \frac{\text{Baseline} - P(1_a, 1_b)|_{I_{overlap}=1}}{\text{Baseline}} = -\frac{2\text{Re}\{\tilde{t}_1 \tilde{t}_4 \tilde{t}_2^* \tilde{t}_3^*\}}{|\tilde{t}_2|^2 |\tilde{t}_3|^2 + |\tilde{t}_1|^2 |\tilde{t}_4|^2} \qquad (6)$$

Coalescence and anti-coalescence correspond to positive and negative values respectively in Eq. (6). For a 50/50 lossless beam splitter with $|\tilde{t}_1|^2 = |\tilde{t}_2|^2 = |\tilde{t}_3|^2 = |\tilde{t}_4|^2 = 0.5$ and transmission phase equal to $\pi$, perfect 100% colescence is expected [1]. Figs. 2(a) and (b) depict the calculated Hong-Ou-Mandel (anti-) coalescence percentages occurring in the output of the structure A, as a function of the GeTe nano-strip filling ratio and wavelength, for GeTe in crystalline and amorphous phases. The filling ratio is the ratio of the GeTe strip width to the width of the unit cell. The graphs show the dispersion properties from 770 nm to 900 nm. At the design wavelength of 810 nm one can identify two regions, signified by red dashed circles, in which the two-photon interference switches sign from large negative values in (a) to large positive values in (b). For the GeTe nano-strip width of 285 nm in the unit cell length of 450 nm, corresponding to a filling ratio of 0.634 (the center of dashed red circles), in the phase transition from crystalline to amorphous the (anti-) coalescence value switches from -97.7% to 75.48%. An anti-coalescence of -97.7% is an almost perfect peak in nonclassical HOM two-photon interference and 75.48% denotes a relatively large dip in coincidence measurements.

Figs. 2(c) and (d) are vertical cross sections of the two-dimensional plots in (a) and (b) illustrating the dispersion characteristics of the response in each phase. The red arrows indicate the design wavelength of 810 nm in each plot. Additionally, one can see a rather broadband performance of the metasurface, especially in the two-dimensional plot of the response in the crystalline phase.

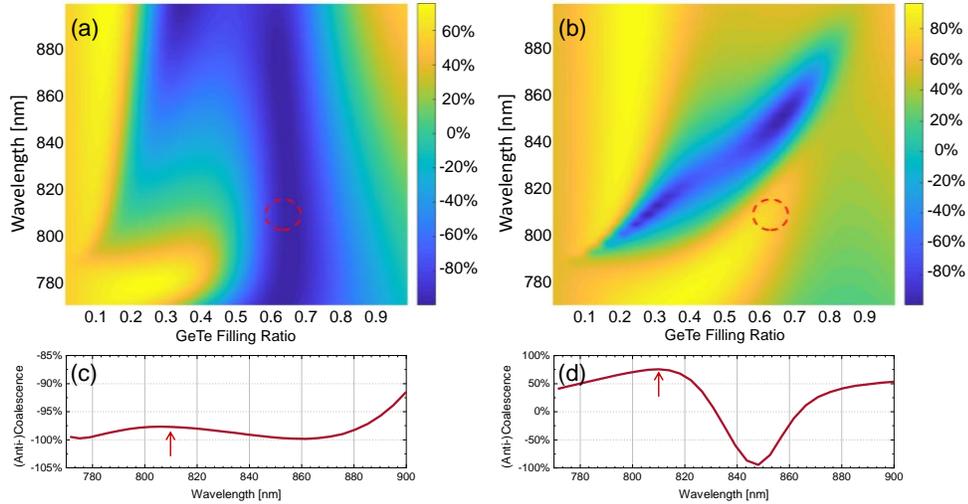

Fig. 2. Hong-Ou-Mandel anti-coalescence (negative) and coalescence (positive) percentages (see Eq.(6)) for structure A shown in Fig. 1(a) as a function of the GeTe nano-strip filling ratio and wavelength for GeTe in (a) crystalline and (b) amorphous phases. (c) Cross section of part (a) at the design wavelength of 810 nm. (d) Cross section of part (b) at the design wavelength of 810 nm.

By adjusting the layers thicknesses in structure B, such as the thickness for lower $TiO_2$, one can easily alter the operating wavelength of these geometries over a wide wavelength range. This can be seen in Figs. 3 (a) and (b) showing the Hong-Ou-Mandel (anti-) coalescence percentages for these metasurfaces in (a) crystalline and (b) amorphous phases versus the lower $TiO_2$ layer thickness and wavelength. The red dashed circle signifies the designed structure at the operating wavelength of 810 nm. By adjusting the thickness along the diagonal, yellow-colored belt in Fig. 3(b) the operating wavelength may be tuned from at least 770 nm to 900 nm, noting that in the crystalline phase, the dashed red circle, corresponding to the operating wavelength, will always stay in the blue region upon this modification.

In structure B and for the wavelength of 810 nm, the $TiO_2$ layer thickness of 330 nm (the center of dashed red circles) results in switching from -59.42% to 86.09% in (anti-) coalescence upon the change from crystalline to amorphous phase. Similar to structure A, in the amorphous phase, the two photons will bunch at the output of the device, in contrast to the anti-bunching induced in the crystalline phase. The rest of the parameters/thicknesses are kept fixed in this study and are reported in section 2.2. Vertical cross sections of Figs. 3 (a) and (b) are shown in Figs. 3(c) and (d) respectively, illustrating the dispersion properties of the output with the red arrows pointing at 810 nm wavelength. For common narrow bandwidth SPDC sources (typically with signal and idler bandwidth of ≤10 nm) the metasurface design shows a robust non-resonance performance, far beyond the bandwidth requirements imposed by the source.

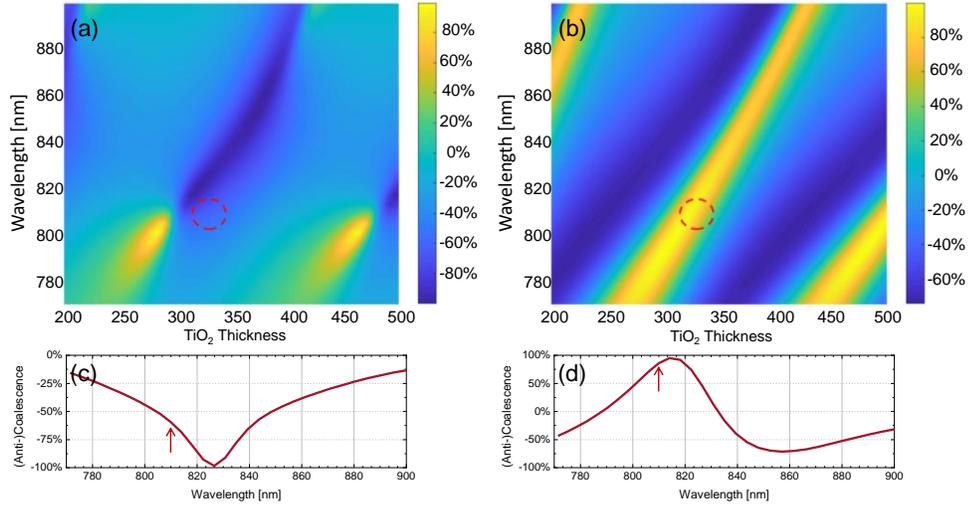

Fig. 3. HOM anti-coalescence (negative) and coalescence percentages (see Eq.(6)) for metasurface B composed of SiO2, GeTe, and TiO2 layers shown in Fig. 1(b) as a function of the lower $TiO_2$ layer thickness and wavelength for GeTe in (a) crystalline and (b) amorphous phase. (c) Cross section of part (a) at the design wavelength of 810 nm. (d) Cross section of part (b) at the design wavelength of 810 nm.

Next, we explore the performance of the devices versus wavelength and the degree of crystallinity of the GeTe. Techniques such as near-field scanning optical microscopy (NSOM) may be employed to study the partially crystallized GeTe films and extract the crystallization fraction using the fact that GeTe material at its two phases exhibits very distinct optical transmission responses [63]. The optical transmission of GeTe films reduces during the transition from amorphous to crystalline phase, in agreement with Fig. 1(d), making it possible to quantify the crystallization factor of the film. This partial phase transition can be captured in the electromagnetic representation of the medium in the form presented in Eq. (7) in which the electric permittivity of GeTe comprises the degree of crystallinity, named $\kappa$, and the permittivites in the crystalline, and amorphous phases, named $\varepsilon_c$ and $\varepsilon_a$ respectively.

$$\varepsilon_{GeTe} = \kappa \varepsilon_c + (1-\kappa)\varepsilon_a \tag{7}$$

Figs. 4(a) and (b) show the values of the (anti-) coalescence percentages for the periodic (a), and layered (b), configurations as a function of GeTe crystallinity and wavelength. Here, for the GeTe/gold periodic geometry (structure A), the structure parameter of GeTe filling ratio is fixed to 0.634, in agreement with the results in Fig. 2. For the layered configuration (structure B), the lower TiO$_2$ layer thickness is fixed to 330 nm, as for the results in Fig. 3. In both 2D graphs, starting from the left side, the degree of crystallinity increases and the PCM tunes from amorphous to crystalline phase. The operating wavelength at both PCM phases are signified with red dashed half circles on the left and right side of the graphs. Figs. 4(c) and (d) depict the cross section of this transition at the operating wavelength. One can observe that the devices' outputs tune smoothly and continuously from positive to negative values in coalescence.

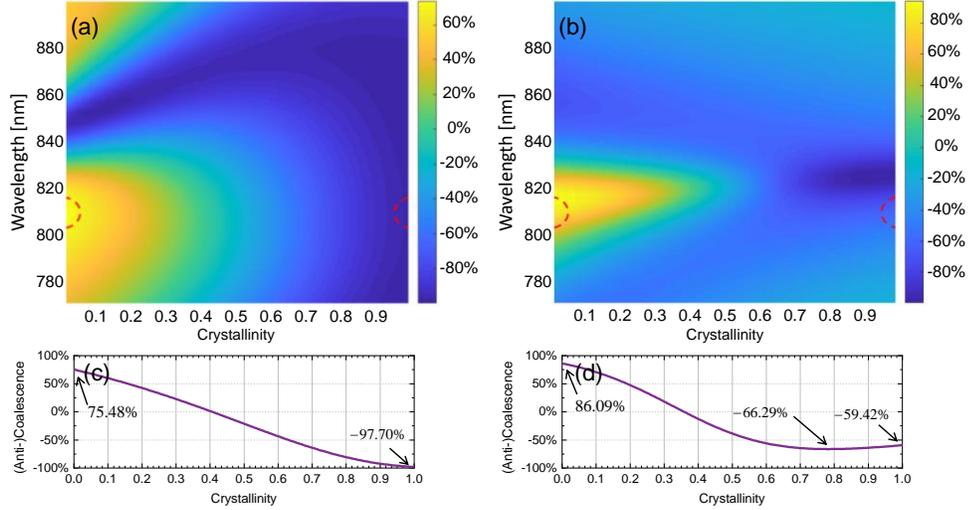

Fig. 4. Hong-Ou-Mandel anti-coalescence (negative) and coalescence percentages (see Eq. (6) for (a) periodic GeTe/gold metasurface structure shown in Fig. 1(a) and (b) layered metasurface composed of SiO2, GeTe, and TiO2 layers shown in Fig. 1(b) as a function of wavelength and degree of crystallinity of GeTe. (c) Cross section of part (a) at the design wavelength of 810nm. (d) Cross section of part (d) at the design wavelength of 810nm.

For the metasurface configuration A (results show in Fig. 4(c)) the output value continuously changes from 75.48% to -97.7% in coalescence during the transition and the tunability is from 86.09% to -59.42% in coalescence for the metasurface geometry B. These smooth transitions also manifest the non-resonant nature of the electromagnetic interactions at these metasurfaces, a factor enabling their moderately high operating bandwidth.

Analyzing temporal aspects of the active-metasurface-enabled tunable fourth-order interference phenomena provides further insights into the tunability. Fig. 5 demonstrates the control over the quantum interference between the two input states of a 2x2 network using the geometries discussed in section 2.2. The normalized second order intensity correlations in the outputs of the network shown in Fig. 1(c) are plotted versus the relative input time delay for two single photon Fock states, one in each input port of the system. This calculated normalized quantity captures the contrast in the rates at which the photons are detected in coincidence versus the time interval between the two input photons. Here, the quantum interference is programmed by setting the crystallinity fraction $\kappa$ of the incorporated PCM (i.e., GeTe).

From the conservation of energy, i.e. $\omega_0 = \omega_a + \omega_b$ in which $\omega_0$ is the angular frequency of the pump photon, the two-photon state vector of Eq. (3) in the down-conversion process is of a single integral form as:

$$|\psi\rangle = \int_0^\infty d\omega_a \phi(\omega_a, \omega_0 - \omega_a) \hat{a}_{in}^\dagger(\omega_a) \hat{b}_{in}^\dagger(\omega_0 - \omega_a) |0\rangle \tag{8}$$

Furthermore, the joint probability for the detection of one photon at the position of detector $D_1$ at time $t$ and another photon at the position of detector $D_2$ at time $t + \tau$ can be found through the following expectation value [64,65]:

$$P_{a,b}(\tau) = K \langle \psi | \hat{E}_{D_1}^{(-)}(t) \hat{E}_{D_2}^{(-)}(t+\tau) \hat{E}_{D_2}^{(+)}(t+\tau) \hat{E}_{D_1}^{(+)}(t) | \psi \rangle \tag{9}$$

This probability, calculated for quantized electromagnetic fields, is essentially the second order correlation function expressing the coincidence counting rate in the two-photon detection approach. The imperfect quantum efficiencies of the detectors are captured in the coefficient $K$ in this formula, and $\hat{E}_{D_1(D_2)}^{(+)}$ and $\hat{E}_{D_1(D_2)}^{(-)}$ are the positive- and negative-frequency parts of the electric field operator at the output port $D_1(D_2)$ (Fig. 1(c)). These operators can be described using the electric field operators at a position prior to the network and time delay segment (which we call $\hat{E}_{a_{0in}(b_{0in})}^{(+)}$ and $\hat{E}_{a_{0in}(b_{0in})}^{(-)}$), the transmission matrix of the system, and the Langevin noise operators. The corresponding negative parts of the electric field operators can be written accordingly noting that these complex operators are mutually adjoint. Throughout the calculation of the joint probability, the contribution of the finite overall time delay associated with the propagation from the point before the time delay to the physical position of the detectors in the experiment ($\tau_1$), does not contribute to the final calculated probability as long as it is symmetric for both arms. In the experiment, this time delay can be finely tuned manually or electronically with respect to the position of the detectors.

$$\begin{aligned}\hat{E}_{D_1}^{(+)}(t) &= \tilde{t}_1 \hat{E}_{a_{0in}}^{(+)}(t - \tau_1) + \tilde{t}_2 \hat{E}_{b_{0in}}^{(+)}(t - \tau_1 + \delta\tau) + \hat{F}_a \\ \hat{E}_{D_2}^{(+)}(t+\tau) &= \tilde{t}_3 \hat{E}_{a_{0in}}^{(+)}(t + \tau - \tau_1 - \delta\tau) + \tilde{t}_4 \hat{E}_{b_{0in}}^{(+)}(t + \tau - \tau_1) + \hat{F}_b \end{aligned} \tag{10}$$

Here, $2\delta\tau$ is the total relative time delay in the HOM geometry introduced symmetrically between the input photons by displacing the 2x2 network from its symmetric position by the distance $c\delta\tau$ where $c$ is the speed of light.

To calculate the positive- and negative-frequency parts of the electric field operators in Eq. (9), we utilize the continuous-mode quantized field operators in the interaction picture by taking the limit of infinite extent for the direction of the propagation of the modes in the input and output ports. Therefore, as an example, the positive-frequency part of the electric field operator can be written as:

$$\hat{E}^{(+)}(z,t) = i \int_0^\infty d\omega \left( \frac{\hbar\omega}{4\pi\varepsilon_0 cA} \right)^{1/2} \hat{a}(\omega) \exp\left[-i\omega(t - z/c)\right] \tag{11}$$

Here, the polarization state is linear and fixed for the input and output ports, thus the field operators may be treated as scalars. We call the 1D Fourier transform of the weight function $\phi(\omega_0/2 + \omega, \omega_0/2 - \omega)$ with respect to the angular frequency $\omega$, function $G(\tau)$, and $g(\tau)$ is the normalized Fourier transform function, i.e. $g(\tau) = G(\tau)/G(0)$. Then, one can show by calculating Eq. (9) for symmetric $g(\tau)$ functions, i.e. $g(-\tau) = g(\tau)$, that the joint probability is equal to:

$$P_{a,b}(\tau) = K|G(0)|^2 \left\{ \begin{array}{l} |\tilde{t}_2|^2|\tilde{t}_3|^2|g(2\delta\tau-\tau)|^2 + |\tilde{t}_1|^2|\tilde{t}_4|^2|g(\tau)|^2 + \\ \tilde{t}_1\tilde{t}_4\tilde{t}_2^*\tilde{t}_3^* g(\tau)g^*(2\delta\tau-\tau) + \tilde{t}_1^*\tilde{t}_4^*\tilde{t}_2\tilde{t}_3 g^*(\tau)g(2\delta\tau-\tau) \end{array} \right\} \tag{12}$$

This time dependent joint probability formula is the generalization for the formalism previously reported in [1] for a lossless and symmetric beam splitter. To link this probability with actual experimental results, in which two detectors and a coincidence counting electronics are employed, with a nonzero and fixed coincidence counting window, the joint probability must be integrated over the coincidence counting window. Given that this window in experimental cases is regularly much longer than the inherent correlation time of the $g(\tau)$ function, which is set in the down-conversion process, the joint probability can be integrated over an infinite time window. This integration results in the total number of coincidences for a real $g(\tau)$ to be proportional to the value $N_{coincidence}$:

$$N_{coincidence} \propto \left[ |\tilde{t}_2|^2|\tilde{t}_3|^2 + |\tilde{t}_1|^2|\tilde{t}_4|^2 + 2\times real\left[\tilde{t}_1\tilde{t}_4\tilde{t}_2^*\tilde{t}_3^*\right] \times \frac{\int_{-\infty}^{\infty} d\tau g(\tau)g(\tau-2\delta\tau)}{\int_{-\infty}^{\infty} d\tau g^2(\tau)} \right] \tag{13}$$

Fig. 5 shows the normalized coincidence counts at the output of the structure A, shown in Fig. 1(a), and the structure B, shown in Fig. 1(b). The two-photon input state in this analysis is the same as the input state applied to the prior studies in this paper. Considering a Gaussian form with bandwidth $\Delta\omega$ for $g(\tau)$ as $e^{-(\Delta\omega\tau)^2/2}$ [1] the Eq. (13) simplifies to the following form and is plotted versus $\tau$.

$$N_{coincidence} \propto \left[ |\tilde{t}_2|^2|\tilde{t}_3|^2 + |\tilde{t}_1|^2|\tilde{t}_4|^2 + 2\times real\left[\tilde{t}_1\tilde{t}_4\tilde{t}_2^*\tilde{t}_3^*\right] \times e^{-(\Delta\omega\delta\tau)^2} \right] \tag{14}$$

As depicted in Fig. 5, through the joule-heating induced phase transition of the GeTe material from amorphous (i.e. $\kappa=0$) to crystalline (i.e. $\kappa=1$), both systems can be tuned monotonically (yet with different slopes) from exhibiting the coalescence effect in which both photons exit the same output port of the network to anti-coalescence in which one photon exits each output port of the network.

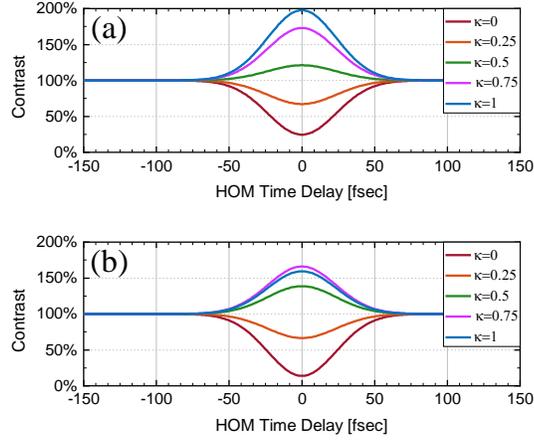

Fig. 5. Nonclassical HOM interference contrast as a function of the relative delay between the two inputs of the 2x2 networks, with single photon Fock states at each input arm of the network for (a) structure A and (b) structure B.

## 3.2 Temperature analysis

In PCM-based technologies the heat required for the phase transition process can be provided optically via lasers or through joule heating schemes. In both of our designs, the phase change process will be realized via integrated joule heating process. In structure A, passing current through the metal strips induces joule heating of the GeTe that fills the spaces between the gold strips. The current is injected through the two gold pads with electric contacts on top of them on the sides of the strips, as shown in Fig. 1(a). In structure B, the TiN heater layer is used as the path for the current to pass through the structure (Fig. 1(b)). In this section we explore the joule heating process leading to phase transition in the non-periodic layered geometry (structure B).

The thermal analyses conducted in this section are done using full 3D FEM (Finite Element Methods) calculations [59]. The layered metasurface arrangement is analyzed in a 3D setup with electric current injected into the structure to induce joule heating. To achieve this objective, the integrated heater and substrate layers are extended, and two gold contact patches are added on top of the heater sheet to simulate realistic electric current ports of the device in joule heating scheme. Realistic values for heat capacity, refractive indices, thermal conductivity, electric conductivity, and material density, acquired from experimental studies, are used in this heat transfer analysis.

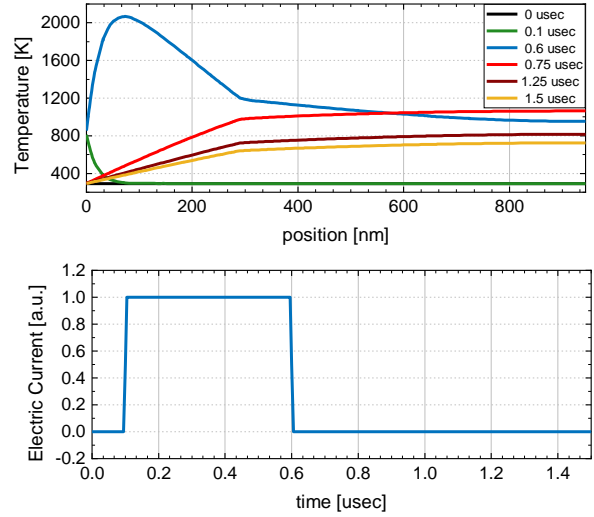

Fig. 6. (a) Temperature profile vs the position inside layered configuration (structure B) at different times during and after applying the electric current pulse. (b) Temporal form of the electric current employed for switching the device from crystalline to amorphous phase. See the text for further explanations.

Fig. 6 displays the simulated induced heating in this structure during the phase transition from crystalline to amorphous. In Fig. 6(a) the temperature profile is plotted versus the position inside the layered structure at different times before, during, and after a 500 ns electrical pulse is injected into the device. The temporal form of the electrical pulse is shown in Fig. 6(b). The lower GeTe layer with a thickness of 21 nm is situated between the positions 290 nm and 311 nm; and the upper GeTe layer with the thickness of 13 nm is situated between the positions 641 nm and 654 nm in this graph.

To examine the device performance comparably to recent experimental efforts, in which matching circuits are often required, we derived the input electric impedance of the geometry [59] and employed the effect of a suitable matching circuit to make a realistic comparison. In this regard similar current levels to those used for making crystallographic phase transitions in [21] are injected into the device. As can be seen in Fig. 6(a), the stimulated local temperature changes at the position of GeTe layers within and after the duration of the pulse fall within inside the required annealing temperature range, as explained in the supplemental document, supporting the occurrence of phase change. See supplemental document for further temperature analysis results.

## 4. Conclusion

In this work two flat adaptive metasurface structures are proposed and studied to control nonclassical two-photon interference effects in free space. GeTe, a PCM with two stable phases, is employed in both designs resulting in structures with zero static power consumption and very distinct states of operation at each phase. In switching from crystalline to amorphous phase the fourth-order interference in the output of the devices monotonically transitions from constructive to destructive interference, i.e., from anti-bunching to bunching of the photons. GeTe based tunable flat-optics-based devices are appealing to the fast-growing field of quantum optics due to their compactness, high switching speed, durability, and power efficiency.

**Funding**

This work was partially supported by the National Science Foundation (NSF) Center for Photonic and Multiscale Nanomaterials (Grant No. DMR 1120923).

**Disclosures**

The authors declare no conflicts of interest.

See Supplement 1 for supporting content